\def\etal{{\it et al.~\/}}

\def\ie{{\it i.e.~\/}}

\def\ltsima{$\; \buildrel < \over \sim \;$}
\def\simlt{\lower.5ex\hbox{\ltsima}}
\def\gtsima{$\; \buildrel > \over \sim \;$}
\def\simgt{\lower.5ex\hbox{\gtsima}}

\documentstyle[11pt,aaspp4,psfig]{article}
\tighten
\singlespace

\lefthead{Marri, Ferrara \& Pozzetti}
\righthead{High-$z$ SN Counts}

\begin{document}

\title{Gravitational Lensing Effects on High Redshift Type II 
Supernova Studies with NGST}

\author{Simone Marri$^{1,4}$, Andrea Ferrara$^{2,3}$, Lucia Pozzetti$^3$}
\affil{
$^1$Dipartimento di Astronomia, Universit\`a di Firenze, \\
50125 Firenze, Italy \\
$^2$ Joint Institute for Laboratory Astrophysics\\ University of Colorado, 
Boulder, CO 80309
\\ E--mail: ferrara@jilau1.colorado.edu\\
$^3$Osservatorio Astrofisico di Arcetri \\ 50125 Firenze, Italy
\\ E--mail: ferrara@arcetri.astro.it, pozzetti@arcetri.astro.it \\
$^4$Max-Planck-Institut fur Astrophysik, \\
85740 Garching bei Munchen, Germany \\
E--mail: marri@mpa-garching.mpg.de \\}
\begin{abstract}
We derive the expected Type II SN differential number counts, 
$N(m)$, and Hubble diagram for SCDM and LCDM cosmological models,
taking into account the effects of gravitational lensing (GL)
produced by the intervening cosmological mass.
The mass distribution of dark matter halos (\ie the lenses) 
is obtained by means of a Monte Carlo method applied to the 
Press-Schechter mass function. The halos are assumed 
to have a NFW density profile, in agreement with 
recent simulations of hierarchical cosmological models. 
Up to $z=15$, the (SCDM, LCDM) models
predict a total number of (857, 3656) SNII/yr in 100 surveyed 
$4'\times 4'$ fields of the {\it Next Generation Space Telescope}.
NGST will be able to reach the peak of 
the $N(m)$ curve, located at $AB \approx 30(31)$ for SCDM(LCDM)
in $J$ and $K$ wavelength bands and detect (75\%, 51\%) of the 
above SN events. This will allow a detailed study of the 
early cosmic star formation history, as traced by SNIIe. 
$N(m)$ is only very mildly affected by the
inclusion of lensing effects. 
In addition, GL introduces a moderate 
uncertainty in the determination of cosmological parameters
from Hubble diagrams, when these are pushed to higher $z$.
For example, for a ``true'' LCDM with ($\Omega_M=
0.4, \Omega_\Lambda=0.6$), without proper account of GL, one would
instead derive ($\Omega_{M}=0.36^{+0.15}_{-0.12},
\Omega_{\Lambda}=0.60^{+0.12}_{-0.24}$). We briefly compare our
results with previous similar work and discuss the limitations of
the model.
\end{abstract}
\keywords{Cosmology: theory -- dark matter -- gravitational lensing --
supernovae -- galaxy evolution -- methods: numerical -- statistical}

\section{Introduction}

The star formation activity in the universe very likely started with 
the formation of the so-called Pop III objects (Couchman \& Rees 1986,
Ciardi \& Ferrara 1997, Haiman \etal 1997, Tegmark \etal 1997, Ferrara
1998) at redshift $z\approx 30$. According to standard hierarchical models
of structure formation, these small (total mass $M \approx
10^6 M_\odot$ or baryonic mass $M_b \approx 10^5 M_\odot$) objects
merge together to form larger units. Assembling massive galaxies such as 
the ones observed today should take a considerable fraction of the
Universe lifetime. Thus, it might be plausible that star formation 
at high redshift ($z\simgt 5$) occurred at a rate limited by the
relatively small amount of baryonic fuel present in early collapsed structures.
As a consequence, these objects are likely to be faint and would probably
escape the detection of even the largest planned instruments. 

However, unless the IMF in Pop III objects is drastically different from the 
local one, some of the stars formed will end their lives as
supernovae (SNe). At peak luminosity, SNe could outshine their 
host protogalaxy by orders of magnitude and likely become the most 
distant observable sources since the QSO redshift distribution shows an 
apparent cutoff beyond $z\approx 4$ (Dunlop \& Peacock 1990).
Detecting high-$z$ SNe would be of primary importance to clarify the role
of Pop IIIs in the reionization and reheating of the universe, and, in general,
to derive the star formation history of the universe and to pose 
constraints on the IMF and chemical enrichment of the universe (Miralda-Escud\'e \& Rees 1997). 

The issue becomes even more interesting if we consider the gravitational
lensing (GL) effects produced by the intervening cosmological mass distribution
on the light emitted by SNe. The flux magnification associated with this
process has been investigated in detail in a previous paper (Marri \& Ferrara
1998, MF) and found to be substantially dependent on  the adopted
cosmological model, thus high-$z$ SNe seem to be perfect tools to constrain 
cosmogonies. In this paper we use the method outlined in MF but revised to include
realistic lens density profiles. 
MF considered GL effects produced by point
lenses;  here we model dark halos/lenses with NFW (Navarro \etal 1997)
universal density profiles derived from numerical simulations. 
Nonetheless, the general numerical methods and principles are
the same as in MF and we refer to that paper for details; indeed 
this more accurate lens modeling improves
the predictive power of our calculations to a level comparable
to rayshooting methods based on N-body simulations, for which an extension
to the high redshift studied here has not yet been possible.
 
We concentrate mostly on Type II SNe (SNII), although it is 
straightforward to
extend our results to include Type I SNe (SNI). The motivation for this choice is
essentially the same as in MF.  SNIa are on average brighter than SNII by
about 1.5 mag; moreover, SNIa are known to be very good standard candles
and, for this reason, they are widely used to determine the geometry of
the universe (Riess \etal 1998; Perlmutter 1998).
However, it is reasonable to expect that SNI at $z\simgt 5$ constitute 
very rare events, 
since they arise from the explosion of C-O white dwarfs triggered by 
accretion from a companion; this requires evolutionary timescales comparable or larger than the Hubble 
time at those redshifts. A different possibility would be represented by SNIb, which
originate from short-lived progenitors: however, they share problems similar to
SNII, \ie they are poorer standard candles and are fainter than SNIa.  
	 
Two different 
cosmological models are considered here: {\it (i)} Standard Cold Dark
Matter (SCDM), with $\Omega_M=1, \Omega_\Lambda=0$,
and {\it (ii)} Lambda Cold Dark Matter (LCDM):  $\Omega_M=0.4,
\Omega_\Lambda=0.6$. 
Power spectra are normalized to give the correct cluster abundance 
at $z=0$ ($\sigma_8 =0.57, 0.95$ for SCDM and LCDM, respectively)
and we adopt the value $h=0.65$ for the adimensional Hubble constant
($H_{0}=65~{\rm Km}~{\rm sec}^{-1} {\rm Mpc}^{-1}$).
We then derive high-$z$ SNII number counts expected when GL           
flux magnification is included for typical parameters and
planned observational capabilities of the {\it Next Generation Space Telescope}
(NGST, Stockman \etal 1998). The present results may serve as a guide 
for future mission operation mode planning.

\section{Cosmic SNII Rate}

To derive    the cosmic rate of SNII as a function of redshift we
begin by calculating the number density of dark matter halos in
the two cosmologies specified above. This can be accomplished by 
using the Press \& Schechter (1974, hereafter PS) formalism; this
technique is widely used in semi-analytical models of galaxy formation
(White \& Frenk 1991, Kauffman 1995, Ciardi \& 
Ferrara 1997, Baugh \etal 1998, Guiderdoni \etal 1998) and it 
has been shown to be in good
agreement with the results from N-body numerical simulations. 
According to such a prescription we can write the 
normalized fraction
of collapsed objects per unit mass at a given redshift as
\begin{equation}
f(M ,z)=\sqrt{\frac{2}{\pi}}
\frac{\delta_{c}(1+z)}{\sigma(M)^{2}}
e^{-\delta_{c}^{2}(1+z)^2/2\sigma(M)^{2}}
\left( -\frac{d\sigma(M)}{dM} \right),
\label{fmz}
\end{equation}
where $\delta_c=1.69$ is the critical overdensity of perturbations
for collapse, and $\sigma(M)$ is the gaussian variance of fluctuations 
on mass scale $M$. 
Next we calculate the star formation rate corresponding
to each collapsed halo. Following Ciardi \& Ferrara (1997) and
Ferrara (1998) we can write the SNII rate per object of total 
mass $M$ as
\begin{equation}
\label{gamma}
\gamma(z)={\nu \Omega_b f_b \over\tau t_{ff}}M\simeq
1.2\times 10^{-7}\Omega_{b,5} f_{b,8}(1+z)_{30}^{3/2} M_6
{\rm ~yr}^{-1},
\end{equation} 
where $(1+z)_{30}=(1+z)/30$, $M_6=M/10^6 M_\odot$. 
We assume a Salpeter IMF with a lower cutoff mass equal to $0.1~M_\odot$, 
according to which one supernova
is produced for each 56~$M_\odot=\nu^{-1}$ of stars formed.
The baryon density parameter is $\Omega_{b}=0.05 \Omega_{b,5}$, of
which a fraction $f_b \approx 0.08 f_{b,8}$ (Abel \etal 1998) is able to cool
and become available to form stars. The halo dark matter 
density is $\rho \simeq 200 \rho_c = 200
[1.88\times 10^{-29} h^2(1+z)^3]$~g~cm$^{-3}$; 
the corresponding free-fall time is $t_{ff}= (4\pi G \rho)^{-1/2}$.
The star formation efficiency $\tau^{-1}=0.6 \%$ is calibrated on the
Milky Way. 

\begin{figure}[t]
\centerline{\psfig{figure=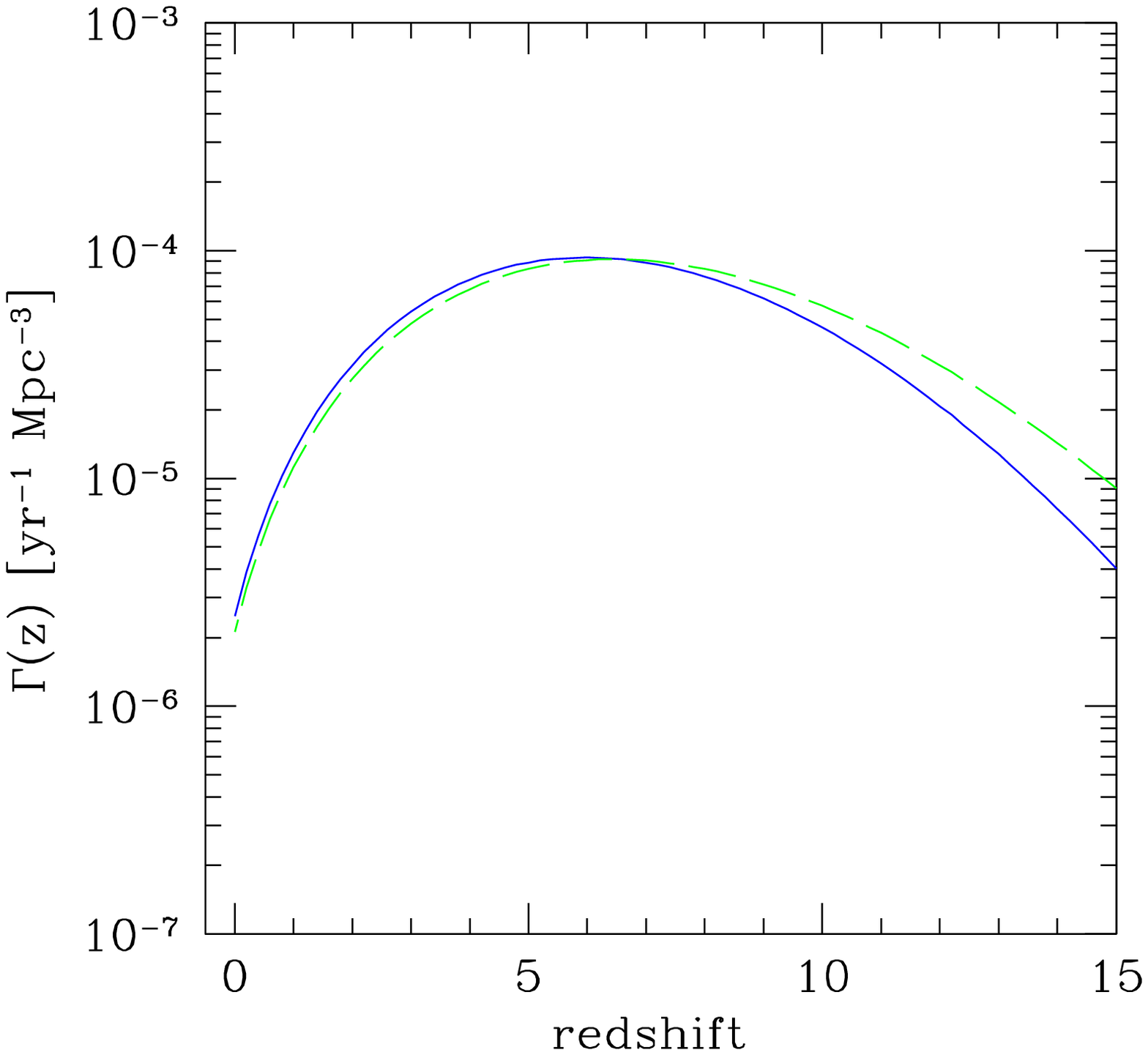}}
\caption{\label{fig1} Comoving rate of Type II SNe as a function 
of redshift for the two different cosmological models: SCDM (solid line), 
LCDM (dashed).
}
\end{figure}

It is well known that the star formation prescription eq. \ref{gamma} 
would lead to a very
early collapse of the baryons - with consequent star formation - 
in small halos, contrary to what is currently believed (Madau \etal
1996). It is therefore necessary to introduce some form
of feedback, most likely due to supernova energy input into the 
interstellar medium of the forming galaxy.
To this aim we follow the standard approach used in semi-analytical 
models (Kauffmann, Guiderdoni \& White 1994; Baugh, Cole \& Frenk 1996,
Baugh \etal 1998). The star formation rate (and consequently the SN rate
which is proportional to that) is weighted by the feedback 
function $\epsilon(M)$, whose expression is
\begin{equation}
\label{fdbk} 
\epsilon(M)={1\over 1+ \epsilon_0\left(M_c/M\right)^{\alpha}}. 
\end{equation} 
The feedback function contains three free parameters: the efficiency,
$\epsilon_0$, the critical mass for feedback, $M_c$, and the power,
$\alpha$, which expresses the dependence of feedback on galactic mass. 
We choose these parameters in the following way. We first fix the value
of $M_c=10^{10} M_\odot$. The motivation for this choice comes from the 
numerical simulations presented by MacLow \& Ferrara (1999), who have
shown that above that threshold galaxies lose only a negligible
mass fraction of gas following moderately powerful starburst episodes. 
Next, we require the calculated star formation rate to best fit the observed 
star formation rate in the universe at redshift $\simlt 2$ (Lilly \etal
1996, Ellis \etal 1996). As considerable uncertainty is present on the
behavior of the cosmic star formation curve at higher $z$ due to the 
alleged presence of extinguishing dust (Rowan-Robinson \etal 1997, Smail \etal 1997), 
such a choice seems to be the most conservative one. 
The physical interpretation of eq. \ref{fdbk}
is that in low mass objects even a relatively small energy input is sufficient
to heat (or expel) the gas, thus partially inhibiting further star formation.
Although reasonable, the validity of this feedback prescription is uncertain
and it should be taken, lacking a better understanding, only as a 
first approach 
to the problem. The derived comoving cosmic SNII rate, $\Gamma(z)$, is shown in 
Fig. \ref{fig1} for the two   cosmological models considered here. 
SCDM and LCDM models predict similar rates and
at $z\simlt 5$, for example, our rates are broadly consistent with the 
SNII rates inferred using the empirical cosmic star formation
history deduced from UV/optical data (Sadat \etal 1998, Madau 
\etal 1998).

\section{Gravitational Lensing  Simulations}

The peak absolute magnitudes of SNII cover a wide range, overlapping 
SNIa at the bright end, but more generally 1.5 mag fainter. In a recent 
study, Patat \etal (1994) conclude that SNII seem to cluster in at 
least three groups, which they classify according
to their B-mag at maximum as {\it Bright} ($\langle M_B \rangle =-18.7$),
{\it Regular} ($\langle M_B \rangle =-16.5$), and {\it Faint} ($\langle M_B
\rangle =-14$), respectively. Note that the Faint class is constituted 
by a single object, \ie SN1987A.
This classification is based on a limited sample (about 40 SNII), 
and therefore a statistical bias cannot be ruled out.
The results of Patat \etal are also compatible with the empirical 
distribution law given by van den Bergh \& McClure
(1994) which we will use in   
what follows. This local SNII luminosity function (LF)  is assumed
not to evolve with redshift.

To obtain a statistical description of the magnification bias
due to GL, we perform rayshooting simulations as described 
in MF. First, we fix a solid angle, $\omega_{m}=$($7.7$~arcmin)$^2$
and thus a cosmic cone in which matter is distributed among halos whose number density
as a function of mass and redshift is obtained via a Monte Carlo method applied
to the PS mass function. 
This procedure is repeated for each of the 50 slices in which the redshift
range $z=1-10$ is subdivided.
We study the propagation of $N_{l}=25\times 10^6$ light rays, 
uniformly covering at $z=0$ a narrower solid angle $\omega_{r}=1.5\times 10^{-6}~
{\rm sr}$, to avoid spurious border effects. The 
light propagation is studied using the common multiple
lens-plane approximation for a thick gravitational lens,
in which the 3D matter distribution is projected onto such planes. 
Finally, light rays are collected within $N_{s}=1.4\times 10^5$ cells 
on each plane, thus obtaining magnification maps 
as a function of source position and redshift. The lower and
upper limits of the lens mass distribution are equal to
$10^{10} M_\odot$ and $10^{15} M_\odot$, respectively;
all other parameters are the same as in MF. 

The mass distribution inside a lens of total mass $M$ at a given
redshift is supposed to follow the
NFW (Navarro \etal 1997) density profile.
The projected surface matter density, as well as the corresponding 
deflection angle, have been calculated by Bartelmann (1996). 
Using the appropriate numerical routine (kindly provided by J. Navarro),
it is straightforward to calculate the relevant parameters for
each lens, once the mass and redshift of the collapsed object are specified. 

Differently to MF, we adopt here a full beam description of light propagation
(Schneider \etal 1992) and, consequently, the average magnification is equal to unity.
Such a scheme requires on each plane a negative uniform
surface mass density, $\Sigma^{-}_{n}<0$, the total (negative)
mass associated with this density being the one given
by the sum over all lenses in the plane. As 
discussed in Schneider \& Weiss (1988), this approach
automatically guarantees flux conservation without using 
the Dyer-Roeder model, whose correctness is often questioned.

\begin{figure}[t]
\hskip 15cm{\psfig{figure=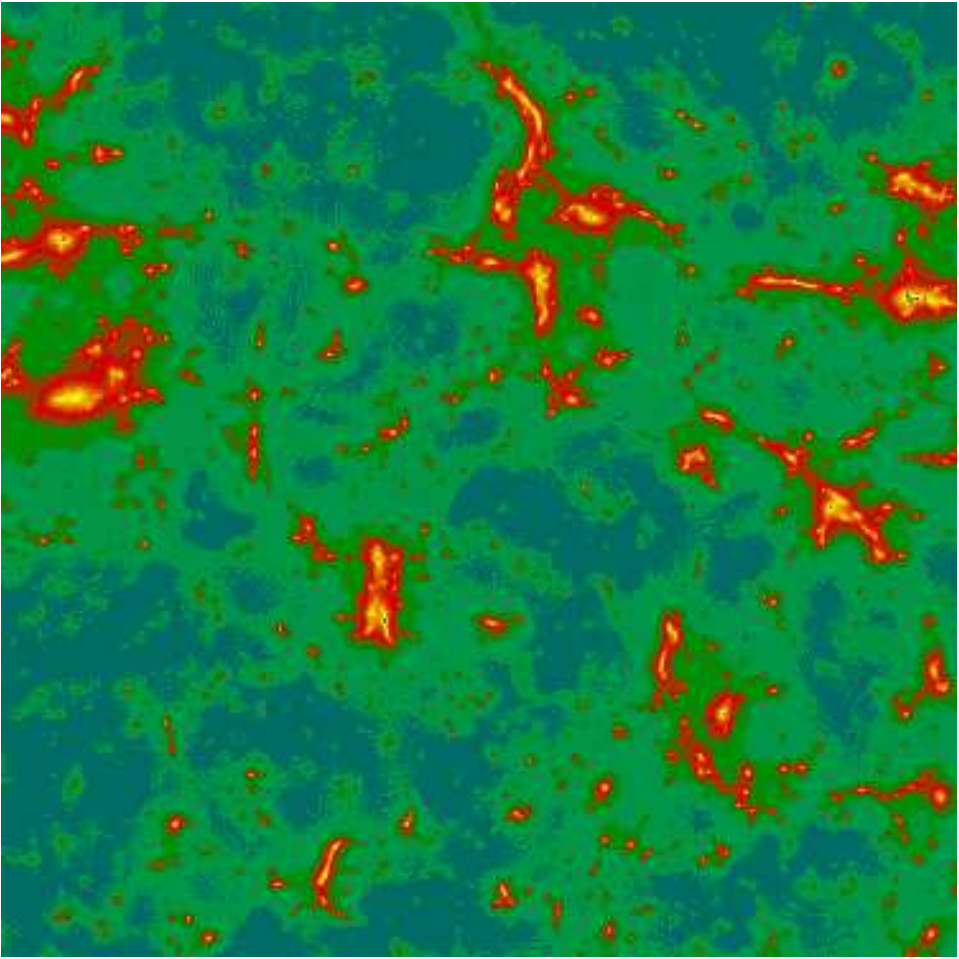}}
\caption{\label{fig1_5} Magnification map for the SCDM model 
(4'$\times$ 4', corresponding to a NGST field) for a source 
located at $z=10$. The magnification range is 0.7-50.
}
\end{figure}

With this prescription 
for the matter distribution, the ray impact parameter
as a function of discretized lens-plane redshift, $\xi_{n}$,
can be recursively calculated from the following expression:
\begin{eqnarray}
\xi_{n+1} = -\frac{(1+z_{n-1})D_{n,n+1}}{(1+z_{n})D_{n-1,n}}\xi_{n-1}
+\frac{D_{n-1,n+1}}{D_{n-1,n}}\xi_{n} \\ \nonumber
-\frac{4G}{c^{2}}D_{n,n+1}\sum_{k=1}^{N_{n}}M^{k}_{n}F^{k}_{n}(|\xi_{n}-\xi_{n}^{k}|) 
\frac{\xi_{n}-\xi_{n}^{k}}{|\xi_{n}-\xi_{n}^{k}|^{2}} \\ \nonumber
-\frac{4\pi G}{c^{2}}D_{n,n+1}\Sigma^{-}_{n}\xi_{n},~~~~~~~~~~~
\end{eqnarray}
where symbols are as in MF, but $D(z)$ is now
the usual Friedmann angular distance and the nondimensional ``form
factor'', $F^{k}_{n}$, depends on the model adopted
for the lens density profile (provided the profile is axially-symmetric),
and it is equal to unity for a point-lens; for a NFW lens the 
$F$-factor can be calculated using the formulae given in Bartelmann (1996). 
As an example, we show the resulting magnification map for the SCDM
model for a source located at redshift 10 in Fig. \ref{fig1_5}. 

\begin{figure}[t]
\centerline{\psfig{figure=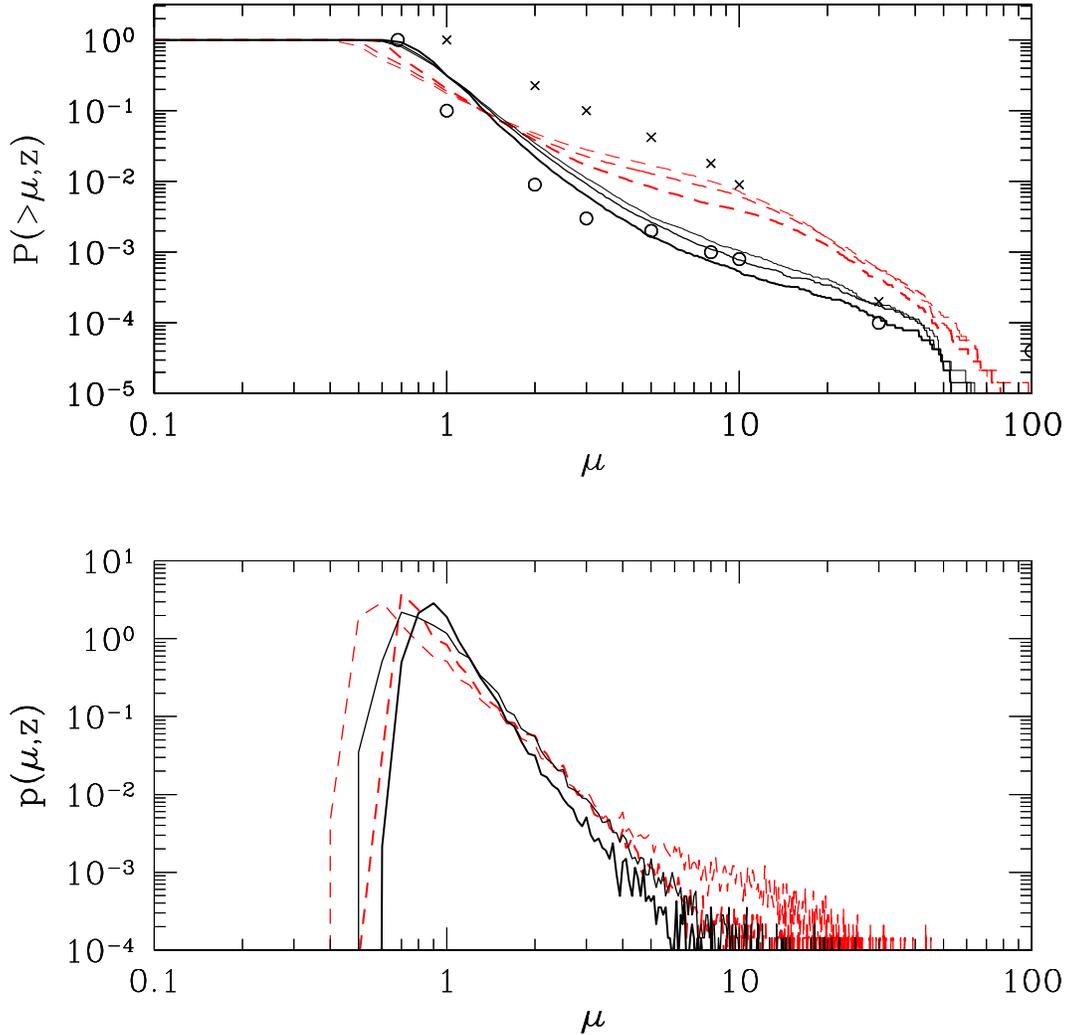,height=15cm}}
\caption{\label{fig2} {\it Top:} Cumulative 
 magnification probability. 
Solid curves are for SCDM, dashed curves refer to LCDM. The
various curves correspond to source redshifts $z=3, 5, 10$
with decreasing thickness, respectively. The circles are the results
of Wambsganss \etal (1998); the crosses are the
results of MF (point lenses). The cosmological model
in both cases is a cluster normalized, SCDM model.
{\it Bottom: } Differential  magnification probability. 
Solid curves are for SCDM, dashed curves refer to LCDM. The
various curves correspond to source redshifts $z=3, 5$
with decreasing thickness, respectively. 
}
\end{figure}

The result of the simulations relevant to the present work are  the          
magnification probabilities at  different redshifts
and for the two cosmological models considered: these are shown in Fig. \ref{fig2}.
The cumulative magnification probability, $P(>\mu)$, which expresses the probability that a
source flux is magnified (or demagnified if $\mu < 1$) more than $\mu$ times, is
shown for $z=3,5,10$. In the lower panel the differential magnification probability is reported for the two redshifts $z=3,5$.
In agreement with previous works (see e.g. Jaroszy\'nski 1992)
we find that LCDM models produce higher magnifications 
than SCDM models. For comparison, we plot the analogous points 
from the CDM  GL simulation of Wambsganss \etal (1998). Also plotted
for the same CDM model are the results of MF: it is clear that
the assumption of point lenses in that work overpredicts the 
magnification with respect to the NFW density profile assumed here.
We note that, as the 
redshift of the source is increased, larger magnifications are allowed: for 
example, at $z=10$ a magnification of 10 times or more has a (cumulative) probability 
of 0.1\% in SCDM and 0.8\% in LCDM models, respectively.
The position of the peak of the differential magnification probability 
shifts toward lower $\mu$ values as $z$ increases and becomes more
flatter in the vicinity of that maximum.
De-magnification also becomes more pronounced, 
and the net effect is an increase in the flux dynamic range.     
Although our simulations explicitly cover the redshift range $0-10$,
the behavior of $P(>\mu)$ shows that the magnification saturates
well before that epoch (see also MF); we can then confidently extrapolate 
the curve even beyond redshift $10$. The following results
are obtained assuming an upper redshift limit $z=15$, 
as also clear from an inspection of Fig. \ref{fig1}.

In order to quantify magnification effects on the  
observed fluxes, we need to perform the following 
steps: 
({\it i}) calculate the total number of SNII by integrating $\Gamma(z)$
in a given redshift interval $\delta z$ around $z_i$ 
and in a certain cosmic solid angle $\omega_{obs}$; 
({\it ii}) assign a peak luminosity to each SNII by randomly sampling
the luminosity function; 
({\it iii}) to assign an amplification to each SN we first check 
for multiple image events. To this aim it is necessary to identify 
the light rays coming from the source in the shooting plane. 
This is done on the fly during the simulations for a
large number of cells on every plane. Once the cell
is identified as the source (\ie a SN), we reconstruct the image
configuration (number of images, splitting, amplification ratios)
by means of an adapted version of the  friend-of-friend algorithm 
which isolates individual images and determines the corresponding  
magnifications.  The very high angular resolution of NGST ($\approx 0.03$
arcsec) should be sufficient to resolve all these multiple images,
whose minimum separation is typically 3 times higher (corresponding to
a $M_{lens}\approx 10^{10}M_{\odot}$).
Each SNe is then characterized by a redshift, an 
intrinsic luminosity and a GL magnification from which 
its observed flux can be obtained using the Friedmann luminosity distance.

As in MF, we adopt $\delta z=0.2$ and
we simulate the observation of 100  NGST fields,
$4' \times 4'$ each, \ie $0.44$~deg$^2$, hence giving a total observed solid
angle $\omega_{obs} \approx 100 \times \omega_{r}$.

We suppose that these fields are surveyed for one year;
in practice, one possible search startegy could be as follows.
Each field is surveyed in two colors and revisited within 
1 month to 1 year to find the SN candidates through their variability.
The typical exposures are about $10^3$~s per color. After selection
of good candidates, a third epoch observation in three colors, with
roughly the same exposure time, is necessary to estimate the redshift
and constrain the light curve. Hence, a total of about 7000 seconds
per field are necessary. For about 100 fields, this observational
time corresponds to a few percent of the total NGST observation time.
In reality, this overestimates the allocated time for the project as
the high-$z$ SN search are also a by-product of the deep galaxy and gravitational
lensing surveys. A similar method, as outlined above, has been applied to
the different individual frames of the HDF-N by Mannucci \& Ferrara (1999)
to discover a $z\approx 1$ SNIb. 

The number of SNII peaks between $5 \le z \le 10$, where
about 1400 SNII occur in the cosmic volume and surveillance time considered.
At even higher redshift (our computations stop at $z=15$)
the total number of SNII events decreases to about 300; 
The two cosmological models (SCDM, LCDM)
predict a total number of (857, 3656) SNII/yr in 100  NGST fields.
The LCDM model offers the best perspectives for SNII detection given
the higher magnification probability, and larger cosmic volume.

\section{Results}

The previous method allows us to build a synthetic sample of SNIIe, 
each of them characterized
by a given luminosity, distance and gravitational amplification. 
This sample can be used for several cosmological applications.
The most natural one consists in the prediction of the number counts
of high-$z$ SNIIe as a function of wavelength bands, 
flux and cosmological model. Detecting these sources would 
enable the study of the early phases ($z \simlt 15$ for our specific
study) of star formation in the universe, as traced by SNII.
We will see in the following that the number counts are very 
weakly affected by the inclusion of GL effects.
However, the number counts turn out
to be rather sensitive to the cosmological models, mostly because
of their different geometry. 
In addition, GL is shown below to produce considerable effects on the
Hubble diagram, commonly used to constrain the geometry
of the universe. We study this effect in the second part of this Section.
This implies that the corresponding uncertainty 
in the cosmological parameter determination should 
be carefully treated and removed. 

\subsection{SNII Number Counts}

In the following we compute the expected SNII number counts 
for our simulated sample and assess the importance of GL 
for such a measurement.
Given the simulated SNII redshift and luminosity distribution 
for the two   cosmological models,
for each SNII we have calculated the apparent AB magnitude  
in four wavelength bands (JKLM), which should cover the 
bandpass of NGST (currently $1-5$~$\mu$m). 	
Assuming a black-body spectrum, $B_\nu(T)$, at 
a temperature $T=25000$~K approximately $15(1+z)$ days after the explosion 
(Kirshner 1990, Woosley \& Weaver 1986), negative 
$K$--corrections ($\simlt -4$ for $z\simgt 4$) allow the detection of SNII 
in the above    bands at high-$z$.
We have taken into account absorption by IGM (Madau 1995);
however, its relevance is very limited as, among the NGST
bands, only the $J$ band at $z\ga 9$ is weakly affected.

\begin{figure}[t]
\centerline{\psfig{figure=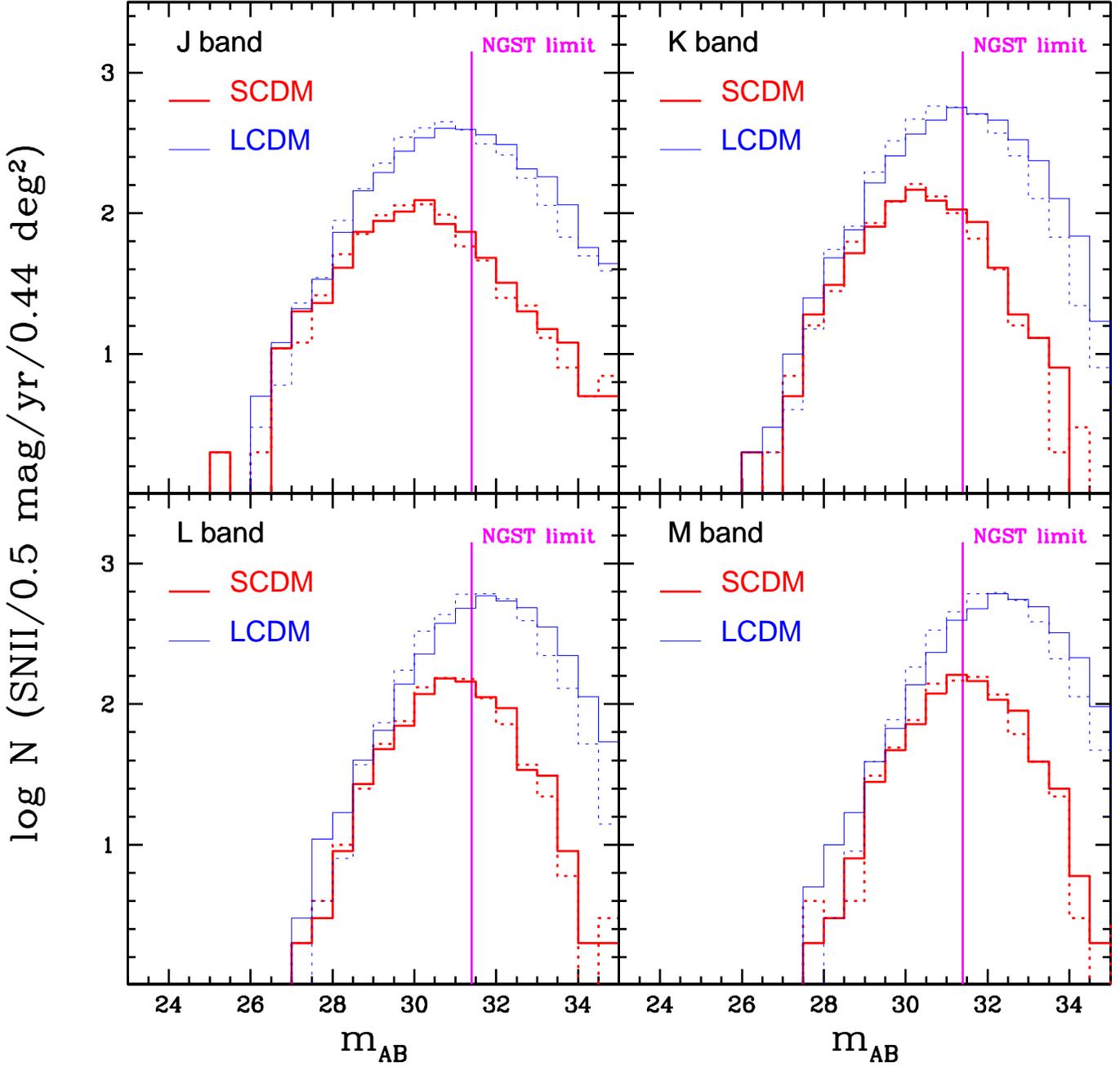}}
\caption{\label{fig4} Differential number counts for the two cosmological 
models
considered ({\it thick solid line}: SCDM, {\it thin solid line}: LCDM) 
as a function of apparent AB magnitude in $J$,$K$,$L$,$M$ bands.
Also shown is the NGST limiting magnitude.
{\it Dashed} curves neglect lensing magnification.
}
\end{figure}

Fig. \ref{fig4} shows the differential SNII counts 
[0.5~mag/ yr/ 0.44 deg$^2$] as a function of AB magnitude. 
The four panels contain the curves for the SCDM and LCDM models and for 
$J$, $K$, $L$, and $M$ bands, both including or neglecting
the effects of GL.                    
For comparison,
we plotted the NGST magnitude limit $AB=31.4$ (vertical line). 
This is calculated by assuming a constant limiting flux 
${\cal F}_{NGST}=10$~nJy in the wavelength range $1-5 
\mu$m (\ie $J-M$ bands). This can be achieved, for a
8-m (10-m) mirror size and a S/N=5, in about $2.6\times 10^4$~s 
($1.1\times 10^4$~s)\footnote{This result has been obtained using
the NGST Exposure Time Calculator, available at http://augusta.stsci.edu}.  
Thus, NGST should be able to reach the peak of expected SNII count 
distribution, which is located at $AB\approx 30-31$ for SCDM and
$AB\approx 31-32$ for LCDM (depending on the wavelength band). 
The differences among the various bands are not particularly pronounced,
although  $J$ and $K$ bands present a larger number of luminous ($AB \simlt 27$)
sources, and therefore they might be more suitable for the experiment.
Furthermore, we point out that in the  $L$ and $M$ bands NGST,
with the current magnitude limit, will not be able to reach the peak 
of expected SNII count distribution in the LCDM model.

The expected number counts are only very mildly affected by the 
inclusion of lensing effects. As a general rule, the curves for both
models become slightly broader and shifted towards fainter magnitudes
by approximately 0.5 mag when gravitational lensing is taken into account. 
These effects are somewhat more important for the LCDM as a result of
the larger magnification dynamic range in this cosmology. 
This behavior derives from the fact that for both models de-magnification 
is relatively probable, particularly at high redshift, as seen in Fig. 
\ref{fig2}. 

However, a clear difference is seen in the number counts between the two
models. At the NGST detection limit in the $K$-band, the (lensed) $N(m)$
ratio (SCDM:LCDM) is equal to (1:5); moreover NGST should be able to
detect $\approx$ 75\%(51\%) of the predicted SNII by SCDM (LCDM) models in
the same band. Differences between bands are similar to those
outlined above.

\subsection{Hubble diagram}

\begin{figure}[t]
\centerline{\psfig{figure=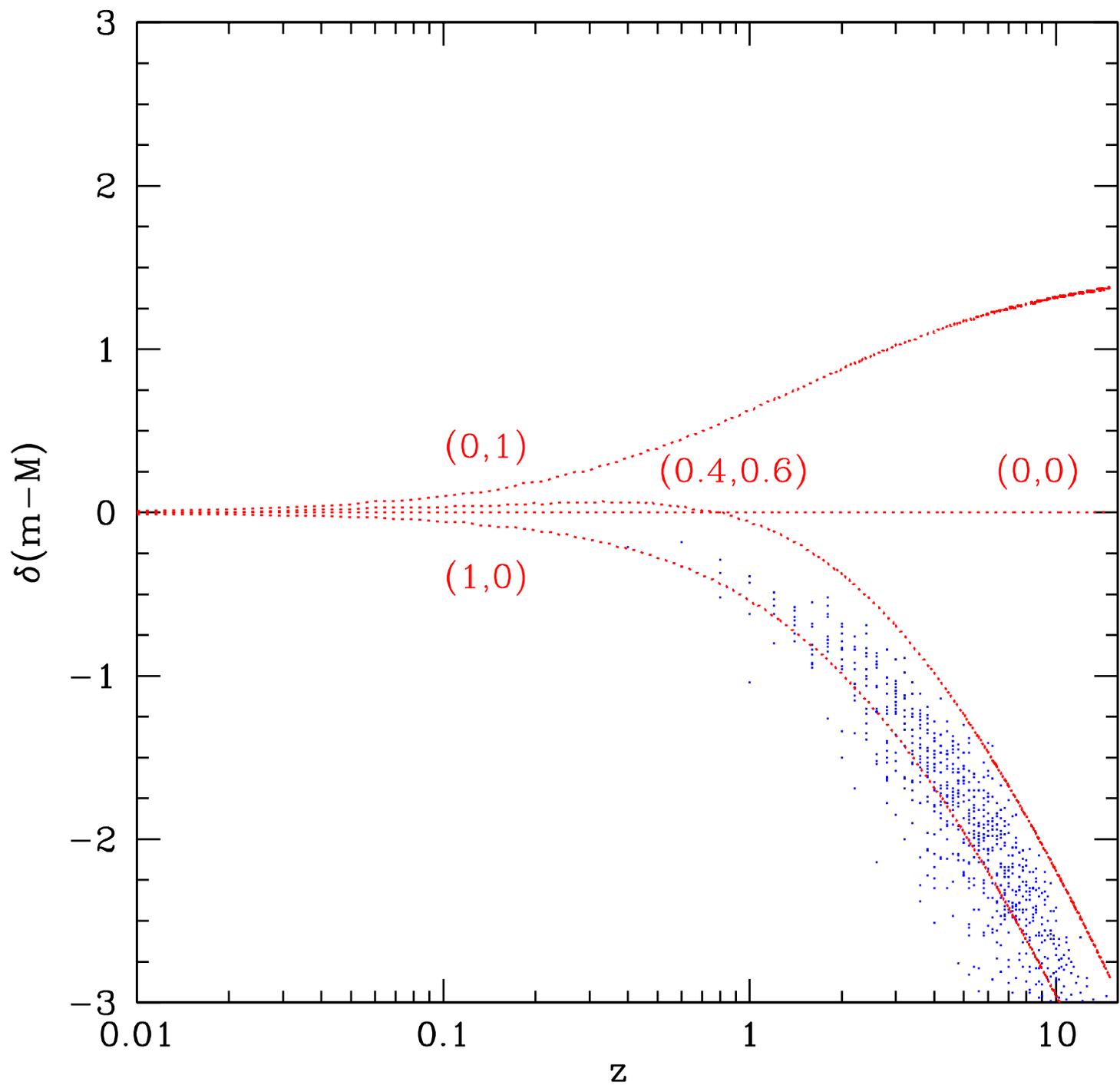}}
\caption{\label{fig5} Hubble diagram for the simulated SNe in the SCDM
model (dots). The dotted lines refer to predictions from different cosmological 
models without lensing effects, values of $(\Omega_{M},\Omega_{\Lambda})$ are indicated. 
}
\end{figure}

\begin{figure}[t]
\centerline{\psfig{figure=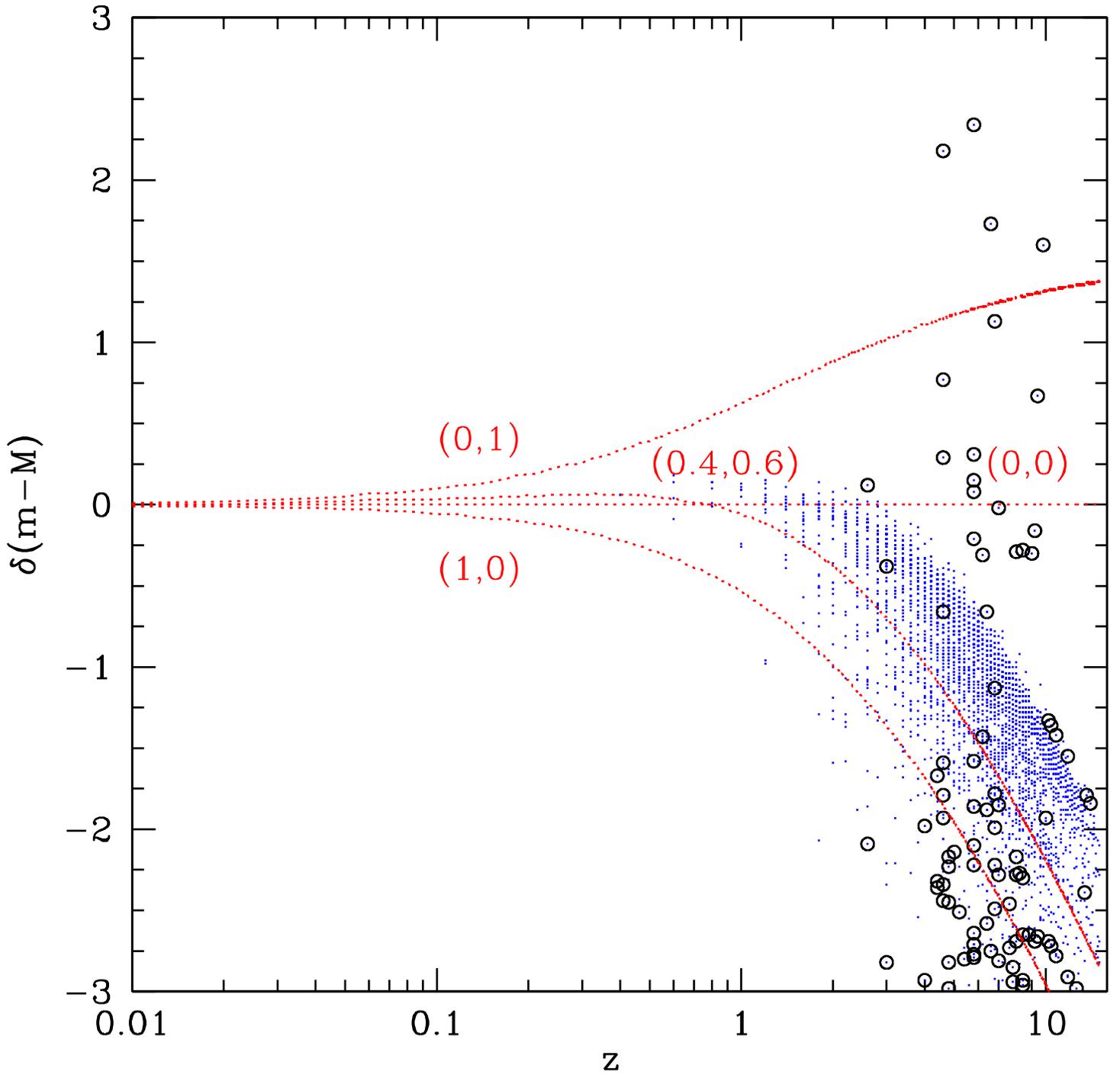}}
\caption{\label{fig6} Same as Fig. \ref{fig5} for simulated SNe in the
LCDM model. Multiple images are displayed with circle dots;
note that both magnified and de-magnified images are present in 
multiple systems.
}
\end{figure}

Although the number counts are weakly affected by GL effects,
the latter must be taken into account when deriving cosmological
parameters from high $z$ SN experiments.
To illustrate this point, we show in Figs. \ref{fig5}-
\ref{fig6} the Hubble diagrams for the SCDM and LCDM models
derived from our sample.
In particular, we have plotted the
distance modulus difference $\delta(m-M)$ between the SNIIe 
in our simulated sample
(shown as points), which includes GL magnification, 
and a specific reference model
with $\Omega_M=0, \Omega_\Lambda=0$. The distance modulus for each
supernova is obtained from the formula $(m-M)_{obs}= 5\log [D_{lum}(z)/\sqrt{\mu(z)}]+25$,
with $\mu(z)$ as derived from our GL simulations discussed above and
$D_{lum}(z)$ expressed in {\rm Mpc}. We assume here that the absolute
magnitude $M$ can be determined within a small error.
Clearly, SNIIe are not as good candles as SNIas commonly
used to construct the Hubble diagram at lower redshifts. For the reasons
already outlined in the Introduction, though, when exploring 
the more distant universe, only SNe whose progenitors were massive
stars might be found, due to the small age of the universe. 
In addition, our understanding of the physics behind these
phenomena is rapidly improving and various methods have been
proposed and successfully tested in the local universe to
derive the absolute magnitude of SNII once the light curve and/or
their spectrum is known. These methods are reviewed in MF.
Even allowing for a persisiting error in their absolute
magnitudes, this spread should have a statistical nature, as opposed
to the systematic one introduced by GL. 
Therefore, an accurate statistical analysis should
be able to disentangle the two effects.  

For sake of simplicity we set $h=0.65$ in the following
discussion and we limit the parameter space to open and flat 
cosmologies with $\Omega_{k}=1-\Omega_{\Lambda}-\Omega_{M} < 0.7$,
$0< \Omega_{\Lambda} <1$ and $0< \Omega_{M} <1$.
The luminosity distance $D_{lum}(z)$ is calculated for a Friedmann model
with $\Omega_{M}$ and $\Omega_{\Lambda}$ values appropriate for the two
cosmological models adopted here, \ie SCDM and LCDM (Fig. \ref{fig5}
and Fig. \ref{fig6}, respectively). For comparison, we have also plotted
the same quantity for the model $\Omega_{M}=0, \Omega_{\Lambda}=1$.
The dotted curves in Figs. \ref{fig5}-\ref{fig6} are thus the distance
moduli which would be observed in absence of any lensing effect
in the cosmological models defined by
the values of $(\Omega_{M},\Omega_{\Lambda})$ indicated above. 
Any source of non systematic errors (as for example the uncertainty 
on the SNII absolute magnitudes) would introduce a quasi-symmetric 
spread of the data around those curves.

The SNe in the sample (we allow for a distinction of multiple images,
that are counted separately and indicated by circled dots in the
Figures; note that high source magnification are associated with 
both amplified and deamplified multiple images) show a peculiar 
spread introduced by magnification/de-magnification effects. 
This spread is more pronounced
for LCDM models, due to the flatter $P(>\mu)$ distribution
obtained for this cosmology (Fig. \ref{fig2}). As a result, the location
of the simulated data points does not lie exactly on the line
corresponding to the LCDM model from which they have been derived;
moreover, the points are not symmetrically distributed around it.

Stated alternatively, if the simulated points would correspond
to real data, the determination of the true cosmological
parameters might be affected by GL effects.
In order to properly estimate the effect we
let $(\Omega_{M},\Omega_{\Lambda})$ to vary and
determine their best fit value by means
of quadratic differences minimization of fluxes:
\begin{equation}
S_{F}=\sum \left[ \frac{F_{i}}{L_{i}} - \frac{1}{4\pi D_{lum}^{2}} \right] ^{2},
\end{equation}
where $F_i$ and $L_i$ are the i-th source flux and luminosity, respectively.
We neglect the intrinsic error on $L_i$, assuming that it could be minimized
as discussed above. Once the best model is found, we look again
for values of $(\Omega_{M},\Omega_{\Lambda})$ which have a quadratic deviation
only $10\%$ larger than the best fit model: $S_{F}<1.1S_{F}^{min}$.
Thus, for the SCDM model we find:
\begin{eqnarray}
\Omega_{M}=0.97^{+0.03}_{-0.18}          \\
\Omega_{\Lambda}=0.01^{+0.03}_{-0.01} \nonumber
\end{eqnarray}
and for the LCDM model:
\begin{eqnarray}
\Omega_{M}=0.36^{+0.15}_{-0.12}  \\
\Omega_{\Lambda}=0.60^{+0.12}_{-0.24} \nonumber
\end{eqnarray}

Thus, we conclude that GL has a moderate impact on  
the determination of cosmological parameters, the largest error being
of order of 40\%.
As a caveat, we remark that if the same best fit method would have been
instead applied directly to the modulus differences rather than on the
fluxes, \ie minimizing the quantity 
\begin{equation}
S_{m}=\sum \left[ \delta(m-M)_{i}-\delta(m-M)_{model} \right] ^{2},
\end{equation}
we would have deduced a much larger, albeit unphysical, discrepancy.
For example, for the LCDM model we would have obtained 
$(\Omega_{M},\Omega_{\Lambda}) \cong (0.3,0.5)$.

Although relatively small, this error due to gravitational magnification
might be relevant for future experiments 
aiming at determining
the geometry of the universe with high $z$ SNe, as if on the one hand
increasing the SN distance would allow for a better discrimination among
different models, on the other hand blurring of the data due to GL
will jeopardize this effort, unless proper account of the effect is
taken.
Models such as the one presented here could be helpful to improve
the fitting procedure, particularly for the most distant SNe
where GL effects are stronger. 


\section{Comments}

A great deal of both theoretical and experimental work 
has recently been devoted to the determination of cosmological
parameters using SNIa at moderate redshifts ($z \simeq  1$). 
As stated above, SNII are likely to be among the most luminous
(and perhaps the only visible) sources in the very distant ($z \approx 
10$) universe. On the other hand, these early epochs could hold
the key for the understanding of the formation of the first objects,
reionization and metal enrichment of the universe. Also, the 
degeneracy among different cosmological models can
be reduced only by extending the current studies to higher 
redshifts. For these reasons, it seems necessary to find 
ways to extract most information from these sources. 

Given that techniques like the ones presented here have been applied
to low redshift SNIe to assess the impact of GL, it seems instructive
to briefly compare our results with those obtained in those works.

Using a numerical method based on geodetic integration, Holz
(1998) calculates the magnification probability and
is then able to "correct" the SNIa Hubble diagram for the effects
of GL. He argues that lensing
systematically skews the apparent brightness distribution of 
supernovae with respect to the filled beam value of the luminosity 
distance. 
It is worth noting that he adopts two different prescriptions
for the dark matter lenses: a constant distribution of halo-like 
objects and a compact object distribution. As was recognized 
also by Metcalf \& Silk (1999), the two distributions yield rather 
different lensing probabilities. In comparison to that work,
our matter distribution is self-consistently derived from CDM 
cosmological models. This appears to be more appropriate to
study the effects arising at redshift greater than unity,
where, despite the common normalization to cluster, 
cosmological models start to differenciate concerning their 
predictions on structure formation. 

Wambsganss \etal (1997), use   a matter distribution
derived from a high resolution N-body numerical simulation of
structure formation in a LCDM model.
Our results agree qualitatively with theirs for what concerns
the magnification probability, although we push our GL simulations
up to higher redshift. In addition, we have been able to predict
the number counts, via a semi-analitycal model, and therefore to
build a synthetic sample enabling us to investigate the effect on
an ensemble of sources rather than on a single one.
As we comment in MF98 our method to reconstruct the matter distribution, 
given a comological model, is by large faster than using numerical
simulations, allowing for an efficient inspection of the parameter 
space. The backdraw is that we miss the information on the spatial
ditribution of lenses. 

We conclude pointing out some limitations of the model
which can be possibly overcome by future work.
Substructure in the halos, as for example 
the one caused by gravitational clustering of dark matter 
within a single halo (Moore \etal 1999) or baryonic dissipative structures
like disks in galaxy halos and, especially, galaxies
in clusters, are not taken into account.
A compact nature of dark matter
could produce different magnification probabilities
and microlensing effects. Recent results have shown that 
the latter effect is probably negligible, as 
microlensing induced variations of SN light curves can  
produce measurable effects only for
lens masses $\simlt 10^{-4}M_\odot$ (see e.g. Kolatt \& 
Bartelmann 1998; Porciani \& Madau 1998).

\bigskip
We thank M. Bartelmann, L. King and P. Schneider for useful comments and 
discussions. Part of this work has been supported by a JILA
Visitor Fellowship (AF); LP acknowledges the support of CNAA during the 
development of this project; SM wish to thank H. Mathis for the help
in adapting the friend-of-friend algorithm.

\newpage

\newpage

\newpage

\newpage

\newpage

\newpage

\end{document}